# NEXT-TO-LEADING ORDER DEBYE-SCREENING IN SPONTANEOUSLY BROKEN GAUGE THEORIES*

Anton K. Rebhan

Theory Group, DESY
Notkestr. 85
D-22603 Hamburg, Germany

## 1. SUMMARY

The main rôle of the Debye screening mass in the perturbative treatment of the electro-weak phase transition is the reduction of the cubic term that determines the strength of a first-order transition. In this note I point out that the standard definition of the Debye mass is unphysical. Its next-to-leading order corrections in resummed perturbation theory are gauge dependent generally in nonabelian gauge theories, and even in Abelian theories when in the Higgs phase. A gauge independent definition can be obtained from a gap equation for the propagator rather than the self-energy, which turns out to be perturbatively under control in the Higgs phase, but sensitive to the nonperturbative magnetic mass scale in the symmetric phase of nonabelian theories.

## 2. ABELIAN HIGGS MODEL

In the Abelian Higgs model with Lagrangian

$$\mathcal{L} = -\tfrac{1}{4}F^2 + |D_\mu \Phi|^2 + \lambda v^2 |\Phi|^2 - \lambda |\Phi|^4, \qquad \sqrt{2}\Phi = \varphi + i\chi \qquad (1)$$

and $R_\xi$-gauge fixing term $\mathcal{L}_{g.f.} = -\tfrac{1}{2\xi}(\partial A - \tfrac{\xi}{2}e\varphi\chi)^2$, the leading-order results for the various masses at high temperature read

$$m_L^2 = \tfrac{1}{3}e^2 T^2 + m^2, \qquad m_T^2 = m^2 \equiv e^2\varphi^2, \qquad (2)$$

$$m_\varphi^2 = 3\lambda\varphi^2 - \lambda v^2 + (\tfrac{e^2}{4} + \tfrac{\lambda}{3})T^2, \qquad (3)$$

$$m_\chi^2 = \lambda\varphi^2 - \lambda v^2 + (\tfrac{e^2}{4} + \tfrac{\lambda}{3})T^2 + \xi m^2, \qquad (4)$$

where $m_L$ and $m_T$ are the longitudinal (Debye) and the transverse (magnetic) mass of the photon propagator.[1]

---

* Contributed talk at the NATO Advanced Research Workshop "Electroweak Physics and the Early Universe", 23 – 25 March 1994, Sintra, Portugal

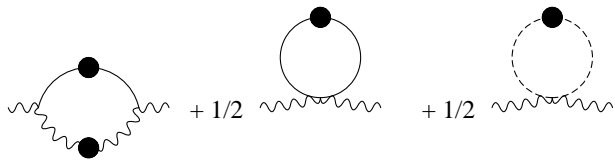

**Figure 1.** Dressed one-loop diagrams for $\Pi_{00}$ in the Abelian Higgs model.

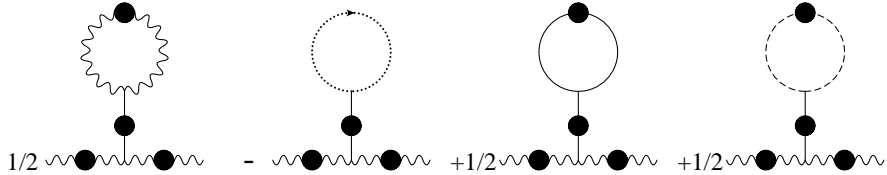

**Figure 2.** Additional dressed one-loop corrections to the longitudinal gauge boson propagator. Wavy, dotted, full, and dashed lines correspond to gauge bosons, Faddeev-Popov ghosts, Higgs and Goldstone particles, resp.; a blob on these lines marks one-loop dressed propagators.

The next-to-leading order result for $\Pi_{00}(0)$, which is usually taken as the definition of the Debye mass squared, is given by the dressed one-loop diagrams of Fig. 1,[1]

$$\Pi_{00}(k_0 = 0, k \to 0) = \tfrac{1}{3}e^2 T^2 + m^2 - \frac{e^2 T}{4\pi}\left(\frac{4m^2}{m_L + m_\varphi} + m_\varphi + m_\chi\right). \qquad (5)$$

Because all higher-order calculation to date have been performed in the Landau gauge, it seems to have gone unnoticed that this definition of the Debye mass is gauge dependent through its dependence on the Goldstone boson mass (4), so that it cannot be the correct one.

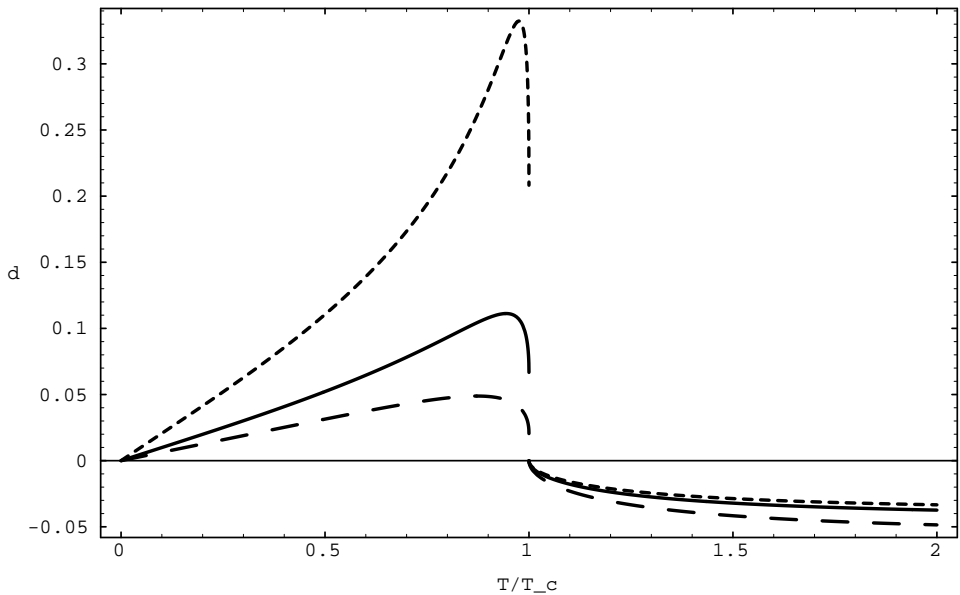

**Figure 3.** $d \equiv \delta m_L/m_L$ over $T/T_c$ in the Abelian Higgs model for $e = 0.3$. The full line is for $\lambda = e^3$; the short-dashed one for $\lambda = e^4$, where the high-T expansion ceases to apply; the long-dashed one for $\lambda = e^2$, where perturbation theory breaks down.

Defining instead[2] the correction to the leading-order Debye mass (2) through the position of the pole of the longitudinal gauge boson propagator at imaginary wave vector $k^2 = -m_L^2$,

$$\delta m_L^2 = \delta \mathcal{D}_{00}^{-1}(k^2 \to -m_L^2), \qquad (6)$$

one indeed is led to a different result. Firstly, the momentum-dependence of the self-energy diagrams has to be taken into account, which is non-trivial in the first diagram of Fig. 1, and secondly, there are additional next-to-leading order corrections in the Higgs phase coming from the reducible diagrams of Fig. 2, which account for the corrections to $V'(\varphi)$. Together this yields

$$\delta m_L^2 = -\frac{e^2 T}{4\pi}\left\{2\frac{m^2}{m_L}\ln\frac{2m_L+m_\varphi}{m_\varphi} + m_\varphi \right. \\ \left. + m_\chi(1 - \frac{2\lambda m^2}{e^2 m_\varphi^2}) - \frac{6\lambda m^2}{e^2 m_\varphi} - \frac{2m^2}{m_\varphi^2}(2m + m_L)\right\}. \qquad (7)$$

Evaluated at the minimum of $V(\varphi)$, the term involving the Goldstone boson mass $m_\chi$ becomes gauge independent. In the symmetric phase one has $m_\chi = m_\varphi$, whereas in the Higgs phase the coefficient of the then gauge dependent $m_\chi$ vanishes.

In Fig. 3, $\delta m_L/m_L$ is given as a function of $T/T_c$ for $e = 0.3$ und some values of $\lambda$. Remarkably, this correction term is discontinuous at $T_c$ even when the phase transition itself were second order. However, perturbation theory breaks down for $T$ very close to $T_c$.

## 3. SU(2) HIGGS MODEL

In the nonabelian case, there is a more complicated gauge dependence introduced additionally by the diagrams of Fig. 4. In $\Pi_{00}(0)$ they contribute

$$\frac{g^2 T}{\pi}\left[-m_T + \xi\frac{m_L^2}{m_L+\sqrt{\xi}m_T}\right] \qquad (8)$$

in the case of the SU(2) Higgs model. The last term did not show up in previous analyses because of their restriction to Landau gauge[3].

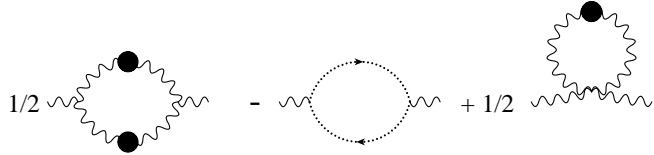

**Figure 4.** Additional dressed one-loop contributions to $\Pi_{00}$ in the nonabelian case.

Evaluating these diagrams at general $k$ however yields

$$\frac{g^2 T}{\pi}\left[-\tfrac{1}{2}m_T - \tfrac{1}{2}m_L + \tfrac{1}{k}(m_L^2 - \tfrac{1}{2}m_T^2 - k^2)\arctan\frac{k}{m_L+m_T}\right. \\ \left. + \left(k^2 + m_L^2\right)\left\{\frac{k^2+m_L^2}{m_T^2 k}\left(\arctan\frac{k}{m_L+\sqrt{\xi}m_T} - \arctan\frac{k}{m_L+m_T}\right) + (\sqrt{\xi}-1)\frac{1}{m_T}\right\}\right], \qquad (9)$$

and the gauge parameter is seen to drop out only at $k^2 = -m_L^2$ as prescribed by the definition (6) for the Debye mass.

Together with the other contributions which are analogous to the Abelian case, the next-to-leading order correction to the SU(2) Debye mass reads

$$\delta m_L^2 = \frac{g^2 T}{\pi}\left[-\tfrac{1}{2}m_L - \tfrac{1}{2}m_T + (m_L - \frac{m_T^2}{4m_L})\ln\frac{2m_L+m_T}{m_T}\right. \\ \left. - \frac{m^2}{8m_L}\ln\frac{2m_L+m_\varphi}{m_\varphi} + \tfrac{1}{8}m_\varphi + \frac{3g^2}{64\lambda}(2m_T + m_L)\right] \qquad (10)$$

in the Higgs phase. In the symmetric phase, however, one encounters a logarithmic singularity due to the vanishing of $m_T$ in perturbation theory. Assuming a nonvanishing magnetic mass $m_T \ll m_L$, one is led to

$$\delta m_L^2 = \frac{g^2 T}{\pi} \left[ m_L \left( \ln \frac{2m_L}{m_T} - \frac{1}{2} \right) - \frac{1}{4} m_\varphi \right]. \tag{11}$$

The sensitivity of (11) to the magnetic mass scale in the symmetric phase means however that perturbation theory is no longer under control. Only the coefficient in front of $\ln(2m_L/m_T) \sim \ln(1/g)$ is reliably calculable. The full result (11) only holds for the (rather crude) assumption of a simple mass term in the transverse propagator.

In Fig. 5, $\delta m_L/m_L$ is plotted for $g = 0.66$, $m_H = m_W$ and $m_{top} = 2m_W$. For $T$ very close to $T_c$, where a (finite) discontinuity in $\delta m_L$ arises, perturbation theory breaks down (as in the Abelian case) because of the vanishing of the Higgs mass. In the symmetric phase a value $m_T = 0.28 g^2 T$ for the hypothetical magnetic screening mass has been adopted, which is consistent with some lattice simulations[4], a recent semiclassical result[5], and also with the result presented at this Workshop by Philipsen[6].

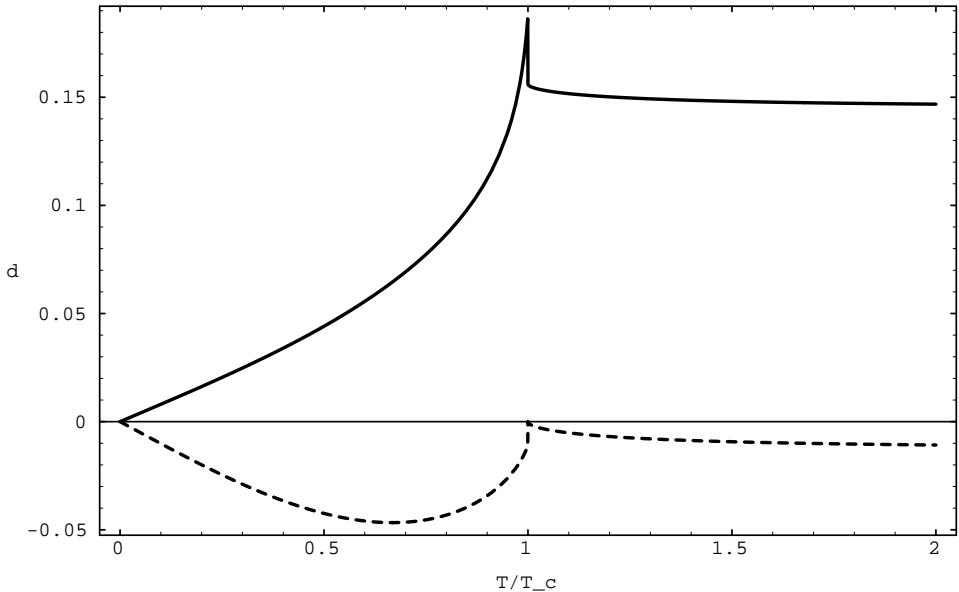

**Figure 5.** $d \equiv \delta m_L/m_L$ over $T/T_c$ in the SU(2) Higgs model for $g = 0.66$ and $m_H = m_W = m_{top}/2$. The result in the symmetric phase ($T > T_c$) depends on the magnetic screening mass, which has been chosen as $0.28 g^2 T$. The dashed line gives the "off-pole" result in Landau gauge[3].